\documentclass[sigconf,screen,authorversion,nonacm]{acmart} 

\usepackage{graphicx}
\usepackage{multirow}
\usepackage{adjustbox}

\usepackage{tikz}
\usetikzlibrary{arrows,shapes, automata}
\usetikzlibrary{calc}
\usetikzlibrary{positioning}
\usetikzlibrary{arrows.meta}
\usetikzlibrary{fit}
\usetikzlibrary{matrix}

\AtBeginDocument{%
  \providecommand\BibTeX{{%
    \normalfont B\kern-0.5em{\scshape i\kern-0.25em b}\kern-0.8em\TeX}}}

\setcopyright{acmcopyright}
\copyrightyear{2022}
\acmYear{2022}
\acmDOI{10.1145/1122445.1122456}

\acmConference[MSR 2022]{MSR '22: Proceedings of the 19th International Conference on Mining Software Repositories}{May 23–24, 2022}{Pittsburgh, PA, USA}
\acmPrice{15.00}
\acmISBN{978-1-4503-XXXX-X/18/06}

\usepackage{marginfix} 

\begin{document}

\title{TSSB-3M: Mining single statement bugs at massive scale}

\author{Cedric Richter}
\affiliation{
  \institution{Carl von Ossietzky University Oldenburg}
  \streetaddress{Ammerländer Heerstraße 114-118}
  \city{Oldenburg}
  \country{Germany}
  \postcode{26129 }
}
\email{cedric.richter@uni-oldenburg.de}

\author{Heike Wehrheim}
\affiliation{
  \institution{Carl von Ossietzky University Oldenburg}
  \streetaddress{Ammerländer Heerstraße 114-118}
  \city{Oldenburg}
  \country{Germany}
  \postcode{26129 }
}
\email{heike.wehrheim@uni-oldenburg.de}

\renewcommand{\shortauthors}{Richter and Wehrheim}
\newcommand{\inlineheadingbf}[1]{\medskip\noindent{\bfseries #1.}}

\begin{abstract}
Single statement bugs are one of the most important
ingredients in the evaluation of modern bug detection and
automatic program repair methods. By affecting only 
a single statement, single statement bugs represent a type of bug often
overlooked by developers, while still being small enough
to be detected and fixed by automatic methods.
With the rise of data-driven automatic repair the availability of single statement bugs 
at the scale of millionth of examples is more important
than ever; not only for testing these methods but also for providing 
sufficient real world examples for training. To provide access to bug fix datasets
of this scale, we are releasing two datasets called \texttt{SSB-9M} and \texttt{TSSB-3M}.
While \texttt{SSB-9M} provides access to a collection of over 9M general single statement bug fixes
from over 500K open source Python projects ,
\texttt{TSSB-3M} focuses on over 3M single statement bugs which can be fixed solely by a single statement change.
To facilitate future research and empirical investigations, we annotated each bug fix with one
of 20 single statement bug (SStuB) patterns typical for Python together with a characterization of the code change as a sequence of AST modifications. 
Our initial investigation shows that at least 40\% of all single statement bug fixes mined fit at least one SStuB pattern, 
and that the majority of 72\% of all bugs can be fixed with the same syntactic modifications as needed for fixing SStuBs.
\end{abstract}




\maketitle

\section{Introduction}
Software bugs, i.e.~unintended behavior introduced by a programmer mistake, come in many
forms ranging from small errors   affecting only a single token
to major design flaws requiring complete program rewrites. Especially 
small bugs in  a single line or statement only can be easily
overlooked by a developer.
For this reason and to relieve the developer from the burden of manually
localizing and repairing small, simple bugs, several
works have explored automatic methods for bug identification~\cite{pradel2020neural, richter2021deepmutants, allamanis2017learning, pradel2018deepbugs, richter2022learning, hellendoorn2019global} and program repair~\cite{chen2019sequencer, tufano2019empirical, vasic2019neural, lutellier2020coconut}.
While traditional methods~\cite{sadowski2015tricorder, ayewah2008using} often rely on hand-crafted rules to identify potential bugs,
recent methods~\cite{pradel2018deepbugs, pradel2020neural, vasic2019neural} started to explore data-driven techniques. Such techniques, however, require access
to huge databases of known bugs for training. Collecting single statement bugs at the required scale is difficult, though, as they only rarely occur in open source projects. Therefore, existing methods often
had to rely on generated {\em pseudo bugs} for producing a sufficient amount of training data. 

In this paper, we address the lack of available training examples (for Python)  by
releasing two new ultralarge collections of single statement bug fixes. 
For this, we mined
over 500K Python open source projects for bug-fixing commits that only change a single statement. 
By following a mining process similar to previous work \cite{karampatsis2020often}, we derived a dataset of more than 9M bug fixes (\texttt{SSB-9M}). 
Since the collected patches do not always fix the bug in isolation (e.g.~the full patch might require further code edits),
we additionally explored a more restrictive definition of 
single statement bug fixes. More precisely, we filtered our dataset
for code changes that fully patch a bug with a single statement change - without modifying source code in other files or statements in the same file at a different location.
This more restrictive process
produced a collection of over 3M "true" single statement bugs (\texttt{TSSB-3M}). 

To compare the distribution of bugs in our datasets with previous such collections like PySStuBs for Python~\cite{kamienski2021pysstubs}, we annotate each bug fix with 
one of 20 simple single statement bug (SStuB) patterns. In total, we find
that around 40\% of all mined bugs can be assigned to at least one pattern. To also explore
the remaining bugs, we additionally annotate each bug with an {\em AST edit script}. Using four edit operation types, a script
describes   how the AST of the buggy code has to be changed to arrive 
at its repaired version. Based on the computed edit script, we find that the majority
of single statement bugs (72\% of all collected bugs) can be fixed with the same edit operation needed
to fix a SStuB. Still, there exists a small percentage (2\% of all collected bugs) which are not even
remotely similar to bugs matching a SStuB pattern and, hence, require special care when
addressed with automatic repairing methods.
We hope that a collection of bug fixes at this scale can not only facilitate future research in data-driven
bug detection and repairing methods, but can also shed light on what types of simple
single statement bugs developers introduce in code and how they fix them.

\vspace*{+2pt}
Our datasets and data generation tool are publicly available:
\begin{center}
\href{https://cedricrupb.github.io/TSSB3M/}{https://cedricrupb.github.io/TSSB3M/}
\end{center}

\section{Methodology}
In the following, we provide a detailed overview over our methodology and
the individual steps necessary for building the datasets.

\subsection{Identifying Appropriate Python Projects}
In order to find suitable repositories for mining single statement bugs we employ
the Libraries.io 1.6~\cite{librariesio2020} package index. 
The package index includes
over 2 million references to Python Git repositories which we employ as a starting point for our mining process.
Although mining on popular projects ensures a well-maintained code base, we intentionally decided against excluding repositories based on their popularity. We believe that all bug fixes
are equally valuable for gaining insights.  Repositories of lower popularity might provide
access to bugs which typically occur during the development,  but would be caught before arriving in a better maintained project.
Furthermore, we also include fork projects since they might provide access to bugs not occurring in the original project. To avoid commit duplication, we later remove all duplicate commits. 
Finally, note that our datasets do not include commits from all projects listed in the index since not all
projects are publicly accessible or contain a single statement commit.

\subsection{Mining Single Line Edits}
To effectively mine single statement bugs at the scale of millionth of repositories,
we start by determining potential bug {\em candidates} with a cheap approximative process.
Instead of identifying single statement bugs directly, we first crawl the repositories
for commits that modify exactly one line per modified file.
During this process, we ignore all changes
to the formatting and comments since they do not alter the program behavior. 
Checking whether a commit contains a single line edit can be performed at a textual level without requiring to parse the complete abstract syntax tree (AST) of the program. In addition, lines in Python generally correspond to single statements by design. Even if a statement is defined over multiple lines, we still
capture single statement modifications that are only performed at a single location. 
However, we do not capture modifications to a single multi-line statement at multiple locations.
To finally determine whether a commit contains a single line edit, we compute the difference
between the program before and after the patch. Similar to the Unix diff algorithm, we consider all file modifications as removing old lines and adding the modified lines. To identify single line changes, we count how many lines have been modified by comparing removed and added lines.
For the comparison, we tokenize each line and compare the respective token sequences.
By employing a tokenizer\footnote{For tokenization, we employ the \texttt{tokenize} package.}, we automatically ignore any changes to the code formatting. In addition, our tokenizer is configured to ignore all changes to comments since they do not alter
the program behavior. 
Finally, we store all commits that modify exactly one line
together with all computed file differences. 

For the mining process, the workload is distributed on a cluster of over 1000 workers. Each worker iteratively clones the assigned
projects and searches through the commit history for single line commits. During this process, we ignore all commits that either add or remove complete files, since we are only interested in single code changes. 
The complete mining process took around two weeks and produced a total of over 66M single line edits from more than 500K git projects.

\subsection{Selecting Single Statement Changes}
After collecting single line edits from thousands of projects, we were interested
in finding all {\em single statement} changes. 
For this, we iterate over all collected
commits, while analyzing the previously stored file difference for single statement changes.
We exclude all commits that (1) are duplicates introduced by fork projects, (2) do not contain a full parsable statement 
or (3) change multiple statements in a single line.
For identifying whether a commit changes a statement, we compute the  AST 
for the file difference. It usually contains changed lines in addition to some context code lines.
Although the code in a file difference is generally not enough to build the full AST,
it is enough for the purpose of single statement change identification. 
We employ a best-effort AST parser\footnote{To parse partial code, we employ the tree-sitter library.} to compute
the two ASTs for the code line before and after the fix. To locate the difference in the AST representation, we perform a simultaneous depth-first search, similar to Kampastsis and Sutton~\cite{karampatsis2020often}, until we find the first node where 
the two ASTs differ. We exclude all commits where the computed AST node is not located inside a statement
or is a root for multiple statements. 

After filtering, we now remain with over 28M single statement changes from over 460K Git projects. 
Note that already by deduplication the size of the dataset nearly halves to over 33M single line edits.
Since we believe that this dataset of single statement changes can be valuable
for future research on the evolution of software projects, we publish it under
the name \texttt{SSC-28M}. However, our main focus is still on the study of real single statement bugs.

\subsection{Identifying True Single Statement Bugs}
In this work, we consider two types of bug-fixing commits: 
{\em Single Statement Bugs}, which are obtained following the procedure of previous works,
and {\em True Single Statement Bugs}, which are determined with a more restrictive but also
more precise selection procedure.
First of all, to determine whether a commit is bug-fixing, we check its commit message
for the occurrence of at least one of the following keywords: ‘error’, ‘bug’, ‘fix’, ‘issue’, ‘mistake’, ‘incorrect’, ‘fault’, ‘defect’, ‘flaw’, and ‘type’. This heuristics was repeatedly shown to be highly precise \cite{karampatsis2020often, tufano2019empirical, chen2019sequencer, kamienski2021pysstubs} (with an accuracy of at least 90\%). In addition, this is a common procedure to filter for bug-fixing
commits when the created datasets are too large to be manually inspected~\cite{karampatsis2020often, tufano2019empirical}.

While this heuristics is effective to identify bug-fixing commits, it assumes that commit message 
and code change are related. However, as shown by prior work~\cite{herzig2013impact}, a significant portion of bug fixing commits are entangled commits and, hence, change more code than it is necessary
to fix a bug. Therefore, as soon as we look at the commit not as a whole, but only a part of it (in our case a single statement change), we risk that the change is unrelated to the bug fix. This motivates our second bug category of True Single Statement Bugs (TSSBs). TSSBs refer to bug fixing commits
that fully patch a bug with exactly one single statement change. This is not only a helpful
property for the evaluation of bug detection methods\footnote{A common assumption in the evaluation of bug detection methods is that the code before the patch is buggy and correct after the patch. This is generally not true for commits belonging to traditional single statement bugs.}, but also avoids entangled commits by definition. In addition, we do not apply commit unrolling~\cite{karampatsis2020often, kamienski2021pysstubs}, which would split a single bug-fixing commit into multiple partial fixes.
Now, after filtering the dataset for single statement bugs and true single statement bugs, we obtain two datasets
\texttt{SSB-9M} and \texttt{TSSB-3M}, which contain over 9M and 3M bug fixes, respectively.

\subsection{Characterizing Bug Patching Edits}

For analyzing the types of bugs collected in our datasets, we employ two ways
for characterizing a bug fix: (1) SStuB patterns and (2) AST Edits.
SStuB patterns~\cite{karampatsis2020often} are used to categorize bug fixes into frequently occurring bug type
categories like changes in the usage of function names or binary operators. To categorize the collected bug fixes, we assign each commit to a unique bug pattern. Similar to \cite{kamienski2021pysstubs}, we assign each bug to the most specific category when it is matched by more than one SStuB pattern. In general, we distinguish 20 typical SStuB patterns as identified for Python and described by previous work~\cite{kamienski2021pysstubs}. If a commit does not fit a bug pattern, we assign it to a generic {\em single statement} pattern or {\em single token} pattern (when the commit only modifies a single token). In total, we found  50\% to 60\% of all instances in our datasets to not fit a single SStuB pattern. 

To better characterize how developers make single statement mistakes, we further analyze the AST edits between the code before and after the code change.
 An AST edit script~\cite{falleri2014gumtree} describes how the AST of the buggy code has to be transformed to arrive at its fixed version. Every edit script consists of four types of AST operations:  \texttt{INSERT} (inserting a new AST node at a given location), \texttt{MOVE} (moving an existing AST node to another position in the same AST), \texttt{UPDATE} (updating the value of a single node) and \texttt{DELETE} (removing a single node).
For computing the AST difference, we employ the same two ASTs employed to filter the datasets and computed from the file difference. Then, we compute for each bug patch an AST edit script by employing a reimplementation of the GumTree algorithm~\cite{falleri2014gumtree}.
By characterizing a bug patch through an AST edit script, we are able to analyze which operations have to be supported by a bug repairing method to fix a given bug. Overall, we found that single statement bugs can be fixed on average with 4.20 AST operations in \texttt{SSB-9M} and 4.52 AST operations in \texttt{TSSB-3M}.

\subsection{Datasets} 
As a result of our mining process, we are releasing three new datasets:
(1) \texttt{TSSB-3M}, a dataset of over 3M isolated single statement bug fixes, (2) \texttt{SSB-9M}, 
a dataset of over 9M general single statement bugs and (3) \texttt{SSC-28M}, a dataset
of over 28M general single statement changes. 
Our intention is to facilitate
future research on large scale datasets in software evolution, data-driven bug detection or repair.
 All collected datasets are novel and were not used in previous studies.
To store millionth of bug fixes, all mined data is stored in a compressed jsonlines\footnote{https://jsonlines.org} format. Every dataset entry contains information about (1) the project and commit
hash of the bug fix, (2) the file difference as a Unix diff between the code before the change and
after and (3) additional analytical results such as the identified SStuB pattern and AST edit script. Because of concerns regarding licensing, we cannot redistribute the complete
code related to a bug fix. However, source code is available in the original projects and
can be referenced via our datasets. All released datasets are publicly available
on Zenodo\footnote{https://doi.org/10.5281/zenodo.5845439} and our tool for generating
the datasets can be found on Github\footnote{https://github.com/cedricrupb/TSSB3M}
and Zenodo\footnote{https://doi.org/10.5281/zenodo.5898547}.
\section{Analysis of Datasets}
Our main contribution is a collection of bugs a magnitude
larger than any existing bug dataset.
Nevertheless, we also  explored the datasets
to gain new insights into developer bugs and their fixes.
Specifically, we were interested in the following two questions:
\begin{description}
  \item[RQ1] Do we find the same distribution of SStuBs in our datasets as in previously collected datasets? 
  \item[RQ2] How different are general single statement bugs from those selected by SStuB patterns?
\end{description}

\begin{table}
\caption{SStub pattern statistics for SSB-9M, TSSB-3M and PySStuBs}
\label{tab:sstub-stats}
\centering
\begin{adjustbox}{max width=\columnwidth}
\begin{tabular}{l c c c c c c c c c}
\toprule
{\bfseries SStuB Pattern} & \multicolumn{2}{c}{\bfseries TSSB-3M} && \multicolumn{2}{c}{\bfseries SSB-9M} && \multicolumn{2}{c}{\bfseries PySStuBs \cite{kamienski2021pysstubs}} \\
\cmidrule{2-3} \cmidrule{5-6} \cmidrule{8-9}
& Count & \% && Count & \% && Count & \%  \\
\midrule
Change Identifier Used  & 237K   & 16 &&  659K & 18 && 9K & 12\\
Change Binary Operand &  174K & 12 && 349K & 9 && 4K & 6\\
Same Function More Args & 150K & 10 && 457K & 12 && 10K & 14\\
Wrong Function Name  & 134K & 9 && 397K & 11 && 9K & 12 \\
Add Function Around Expression & 117K & 8  && 244K & 6 && 6K & 9\\
Change Attribute Used & 104K & 7 && 285K & 8 && 5K & 7 \\
Change Numeric Literal & 97K & 6 && 275K & 7 && 5K & 7\\
More Specific If & 68K & 5 && 121K & 3 && 2K & 3\\
Add Method Call & 60K & 4 && 118K & 3 && 3K & 5\\
Add Elements To Iterable & 57K & 4 && 175K & 5 && 2.5K & 3\\
Same Function Less Args & 50K & 3 && 169K & 4 && 3.4K & 5 \\
Change Boolean Literal & 37K & 3 && 82K & 2 && 1.5K & 2\\
Add Attribute Access & 32K & 2 &&  74K & 2 && 1.5K & 2\\
Change Binary Operator & 29K & 2 && 71K & 2 && <1K & 1\\
Same Function Wrong Caller & 25K & 2 && 46K & 1 && 1.2K & 2 \\
Less Specific If & 22K & 2 && 45K & 1 && <1K & 1\\
Change Keyword Argument Used & 20K & 1 && 59K & 2 && 1.5K & 2\\
Change Unary Operator & 15K & 1 && 23K & <1 && 2K & 3\\
Same Function Swap Args & 8K & <1 && 77K & 2 && <1K & 1\\
Change Constant Type & 6K & <1 && 12K & <1 && 2K & 3\\
\midrule
NoSStuB - Single Statement & 1.15M & - && 3.37M & - && - & -\\
NoSStuB - Single Token &  740K & - && 2.20M & - && - & -\\
\midrule
Total SStubs &1.46M & 100 && 3.74M & 100 && 73K & 100  \\
Total    & 3.34M & - && 9.34M & - &&  - & -\\
\bottomrule
\end{tabular}
\end{adjustbox}
\end{table}

\inlineheadingbf{RQ1} We measure the frequency of SStuBs in
 datasets \texttt{TSSB-3M}, \texttt{SSB-9M} and a previously
explored Python SStuB dataset called PySStuBs \cite{kamienski2021pysstubs}. 
The number of occurrences of each pattern in each dataset is given in Table \ref{tab:sstub-stats}. 
In general, we find that the distribution of SStuBs changes between datasets while the {\em ranking} of pattern frequencies is highly correlated (with a Spearman rank correlation of $0.94$ between \texttt{TSSB-3M} and \texttt{SSB-9M} and $0.86$ between \texttt{TSSB-3M} and PySStuBs). In particular, we observe that patterns  such as {\em Change Identifier Used} and {\em Same Function More Args} are highly frequent in all datasets. 
Interestingly enough, some bug categories such as {\em Change Binary Operand} become more dominant in the \texttt{TSSB-3M} dataset. This could indicate that these types of bugs are harder to identify by a developer and, hence, are more likely to be  fixed after the main development phase in an independent fix.
Finally, we find that {\em only 40\% - 43\% of all  Python single statement bugs} fit a SStuB pattern. While this finding is inline with previous observations on Java SStuBs \cite{karampatsis2020often}, we are interested in the types of bugs occurring in the remaining dataset which is what we study in RQ2. 

\begin{figure}
\centering
\scalebox{0.85}{%
\input{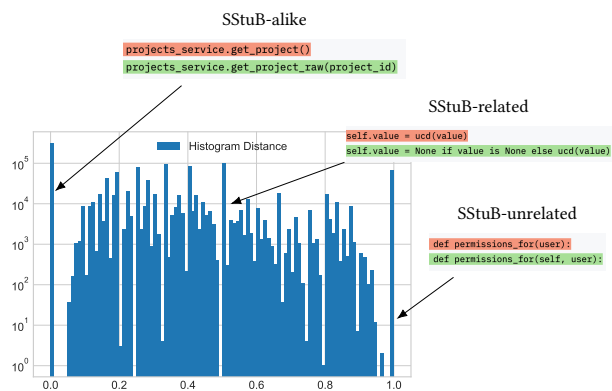}
}
\caption{ 
Similarity of NonSStuBs with SStuBs given as the Jaccard distance between edit operations.
The x-axis corresponds to the binned distance to the most similar SStuB bug   and the y-axis corresponds to observed frequencies. 
}   \label{fig:distance-bin}
\end{figure}

\inlineheadingbf{RQ2} We measure the similarity between single statement bugs that do not fit a SStuB pattern (NonSStuBs) and those that classify as a SStuB in the \texttt{TSSB-3M} dataset. The similarity between bug types
is computed by the Jaccard distance~\cite{jaccard1901etude} between the edit scripts of the respective bug fixes. Therefore, two bug fixes are similar if they share the same edit operations\footnote{Instead of the concrete edit operation that is highly specific to the modified AST we employ the type of an edit operation composed of operation name and types of its arguments. For example, the type of an operation that adds a binary operator to an assignment is \texttt{Insert(binary\_op, assign)}.}. 
For the comparison, we  always look at the minimal distances between NonSStuB and SStuB pairs.  
 Figure~\ref{fig:distance-bin} report on the distribution of computed distances. 
The distance allows us to categorize NonSStuB bugs  
into three classes: SStuB-alike (distance of $0$), SStuB-related (distance of around $0.5$) and SStuB-unrelated (distance of $1$). We find that SStuBs + SStuB-alike bugs populate around 72\% of the dataset and can therefore be repaired by the same edit operations. Another frequent class are SStub-related bugs, which are related to existing SStuB patterns but do not classify as a SStuB. For example, bug fixes that operate on inline \texttt{if}-conditions. While being \texttt{if} statement related, they are not covered by any actual SStuB pattern.
Finally, we find that around 2\% of all bugs (SStuB-unrelated) are completely unrelated to any SStuB pattern. Those types of bugs are often highly related to the specifies of the Python language. For example, we observed that developers commonly forget to add the implicit \texttt{self} argument to a Python method.
The fact that this class of bugs has, however, a low frequency is quite promising for the research on bug detection and repairing methods: {\em Methods that supports the detection and repair of SStuBs are also likely well suited to identify and fix a wide range of the most common single statement bug types.}

\section{Related Work}
Single statement bugs, especially SStuBs, were explored
in previous works, not only as a data source~\cite{karampatsis2020often, kamienski2021pysstubs, beyer2021benchmark} but also for
methods that can detect and prevent this simpler type of bugs~\cite{tufano2019empirical, chen2019sequencer, mashadi2021codebert, lutellier2020coconut}.
Karampatsis and Sutton~\cite{karampatsis2020often} collected 
around 63 thousand Java single statement bugs from 1000 popular projects
that fit at least one of 16 SStuB categories. They found out that around
33\% of all mined bugs can be classified as a SStuB.
Kamienski et al.~\cite{kamienski2021pysstubs} explored similar types of bug fixes for the 1000 most popular Python projects\footnote{Since our dataset mining is more recent and some of the projects mined in PySStuBs might not be public any longer, our datasets are not necessarily supersets of PySStuBs.}
while introducing 7 new SStuB patterns typical for Python. Their collection
of bugs is of similar size with around 73 thousand examples. Similar to Kamienski et al.~we also explored bugs that fit SStuB patterns in Python. However, our mining process allowed us to
go beyond the most popular Python projects, which resulted in a dataset more than 20x larger 
than all existing SStuB collections. In addition and in contrast to previous bug collections,
we also analyze edit operations needed to fix a bug. Therefore, we were able to explore bug
types that are not covered by SStuB patterns.
With the objective of machine-learning based program repair for single line bugs,
Tufano et al.~\cite{tufano2019empirical} collected a set of 787 thousand
bug-fixing single line commits in Java. They used the same heuristics we employed
for identifying bug-fixing code changes. In addition, their method is trained
to translate buggy code lines into its fixed version. Finally, Bader et al.~\cite{bader2019getafix} showed 
that AST edit scripts can be effectively employed for automatic program repair
by predicting bug fixes based on previously seen AST transformations. 
Not only does 
our dataset provide more than 11x more training data (\texttt{SSB-9M}), which has the potential
to improve the performance of existing methods~\cite{halevy2009unreasonable, sun2017revisiting},
but also by annotating each bug fix with an AST script our dataset can directly be employed in various
training setups.

\section{Limitations - Threats to validity}
Even though the employed heuristics for identifying
bug fixing commits has been repeatedly shown to be highly precise,
there is still a chance for false positives in our datasets. To mitigate
this problem, we have designed our mining process to be as precise
as possible by avoiding commit unrolls and by filtering for isolated
bug fixes in \texttt{TSSB-3M}. The latter not only guarantees that the causal relationship between
commit message indicating a bug fix and code change persists, but also avoids the mining of entangled code changes. 
While we mined a large portion of all available open source
projects, we were restricted to projects accessible to the public.
The distribution of simple bugs in closed source projects or projects without version control
might be different. Finally, while the mining process is general enough to be applied to other languages, our datasets are restricted to single statement bugs in Python. The concrete
instantiations and frequency of bugs might vary 
for other programming languages and projects.

\section{Conclusion}
In this work, we explored a new mining process for single statement bugs
which enabled us to crawl through the commits of more than 500K public
git repositories. As a result of our mining process, we are releasing two new
datasets of single statement bug fixes that are a magnitude larger than
any existing collection of bug fixes. More precisely, we introduce \texttt{SSB-9M},
a dataset consisting of over 9M single statement changes in Python, as well as  
 the 3M bug fix dataset \texttt{TSSB-3M}. \texttt{TSSB-3M} guarantees that
every collected bug can be fixed by modifying only a single statement. 
We hope that datasets of this size will facilitate future research in data-driven bug detection and automatic program repair.


\begin{acks}
The authors gratefully acknowledge the funding of this project by computing time provided by the Paderborn Center for Parallel Computing (PC²).
\end{acks}

\bibliographystyle{ACM-Reference-Format}
\bibliography{main}

\clearpage
 
\appendix

\section{Statistics of single statement bugs}
Using the AST edit scripts computed for each single statement bug, we aim to provide
some further insights what is really needed to fix a single statement bug. For this, we answer
the following research questions.

\subsection{How many AST edit operations are typically needed to fix a single statement bug?}
\begin{figure}
\centering
\includegraphics[scale=0.5]{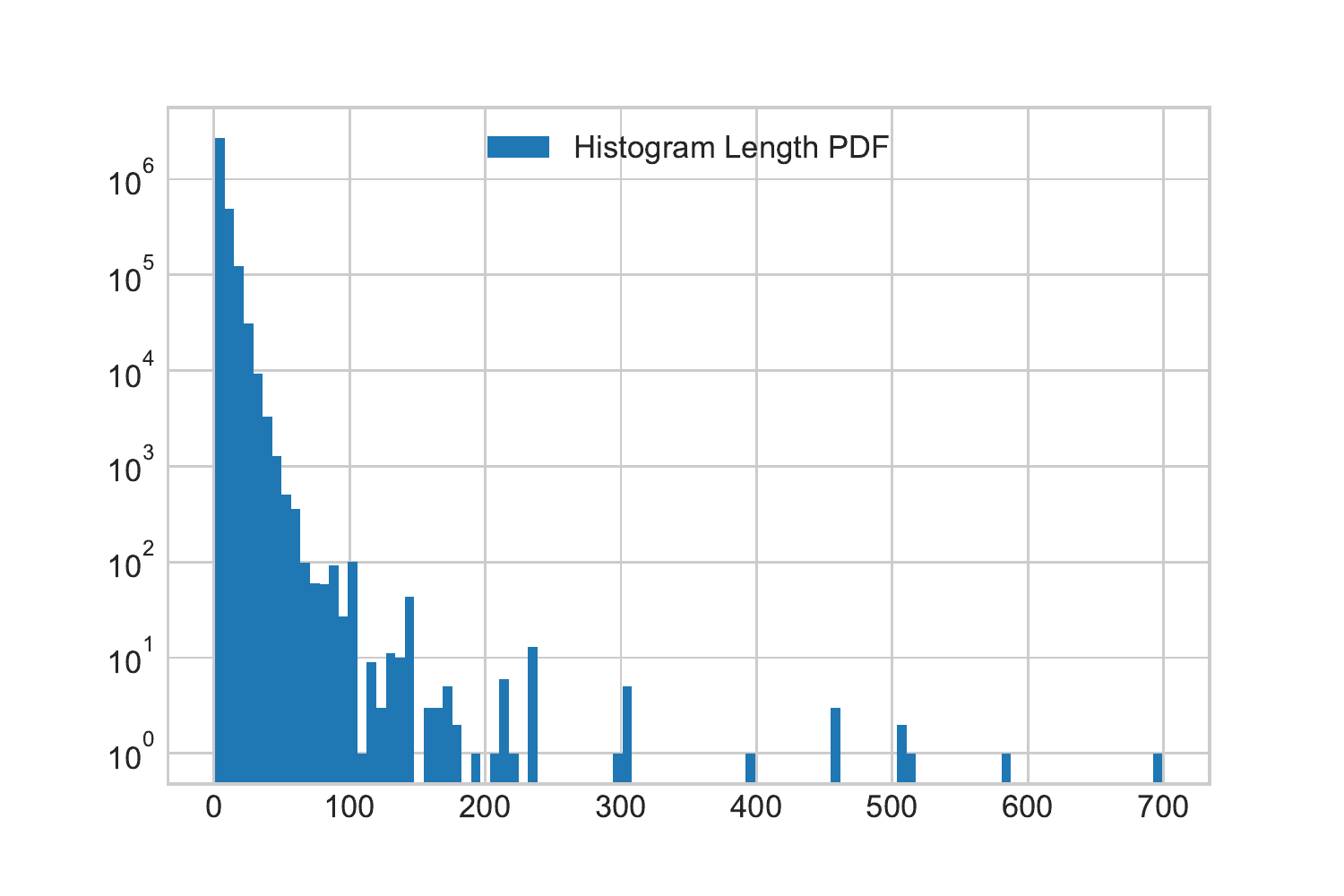}
\caption{Single statement bugs can be typically repaired with only a few AST edit operations.
However, there exists bugs with much larger edit scripts. The largest one employs 3708 operations which we exclude in the histogram for brevity. 
The x-axis corresponds to the binned number of AST operations needed to fix a single statement bug computed on the \texttt{TSSB-3M} dataset. 
}   \label{fig:length-bin}
\end{figure}

Fixing bugs with a short AST edit script is typically easier. These changes only modify simple structures
of the buggy code and do not require to introduce completely new structures. Therefore, we are
interested how the AST edit scripts with varying lengths are distributed. To give an insight, we compute
the length of all AST edit scripts, which is depicted in the histogram in Figure \ref{fig:length-bin}. Overall, we find that
the distribution of AST edit script lengths follow the Zipf law. In other words, single statement bugs that can be fixed with only a few edit operations are much more common than scripts which require a large set of operations. This is promising for the field of automatic repair since researcher can focus on edits to programs that rather small while still covering a large portion of possible single statement bug fixes.

\subsection{What types of AST edit operations are needed to successfully repair a single statement bug?}
\begin{table}
\caption{Typical AST edit operations needed to fix a single statement bug}
\label{tab:ast-ops}
\centering
\begin{adjustbox}{max width=\columnwidth}
\begin{tabular}{c l }
\toprule
{\bfseries Frequency} & {\bfseries Insert\dots (1.55M)} \\
\midrule
441K & identifier into new attribute (\texttt{self.\_} $\rightarrow$ \texttt{self.a})\\
414K & argument list into call (\texttt{f()} $\rightarrow$ \texttt{f(x)}) \\
148K & new boolean operator (\texttt{if ...} $\rightarrow$ \texttt{if ... and ...})  \\
147K & new comparison operator (\texttt{if ...} $\rightarrow$ \texttt{if ... <= ...})  \\
123K & string into argument list (\texttt{f(x)} $\rightarrow$ \texttt{f(x, "arg")}) \\
\midrule
{\bfseries Frequency} & {\bfseries Move\dots (1.25M)} \\
\midrule
172K & identifier to a new attribute (\texttt{self.\_} $\rightarrow$ \texttt{self.a})  \\
109K & identifier to argument list (\texttt{f(x)} $\rightarrow$ \texttt{f(x, y)}) \\
52K   & comparison to boolean operator (\texttt{x < 1} $\rightarrow$ \texttt{x < 1 and ...}) \\ 
48K   & call to argument list (\texttt{f(x)} $\rightarrow$ \texttt{g(f(x))})  \\ 
42K   &  attribute to new attribute (\texttt{a.b} $\rightarrow$ \texttt{a.b.c}) \\ 
\midrule
{\bfseries Frequency} & {\bfseries Update\dots (1.75M)} \\
\midrule
845K & string\\
811K & identifier\\
100K &  integer\\
14K &  floating point number\\
 570 & binary operator \\
 \midrule
{\bfseries Frequency} & {\bfseries Delete\dots (848K)} \\
\midrule
364K & identifier \\
187K & binary operator \\
163K & attribute \\
139K &string \\
110K & call \\
\bottomrule
\end{tabular}
\end{adjustbox}
\end{table}

Automatic program repair (APR) methods often are either tailored to perform certain code change operation
or are more likely to perform certain code transformations. Therefore to guide future research, we investigate
which edit operations are most likely to be needed to fix a single statement bug. Our results are summarized in Table \ref{tab:ast-ops}. We summarize the frequency of each edit type (\texttt{Insert}, \texttt{Move}, \texttt{Update}, \texttt{Delete}) together with top 5 most frequent abstract operation types (see Section 3) in Table \ref{tab:ast-ops}. Before we start interpreting this statistics, note that every type of operation relate to certain feature supported by the APR technique. The most simplest is the \texttt{Update} operation, which only requires to map each token to the same or a new update token in a program. \texttt{Insert} operations require to map a shorter token sequence to a longer one and \texttt{Delete} requires to do the inverse. While a \texttt{Move} operation can be represented as an \texttt{Insert} and \texttt{Delete}, the occurrence of \texttt{Move} operation can motivate the use of a copy mechanism~\cite{gu2016incorporating} which copy some tokens from one location to another.
If we now look at the statistics in Table~\ref{tab:ast-ops}, we will find that operations that modify or extend
the buggy program are much more common than \texttt{Delete} operations. This indicates that a single statement bug fix typically adds something to a program to fix a given bug. Interestingly enough, \texttt{Update} operations have the highest frequency in the dataset. However, we have to note that \texttt{Update} operations are dominated 
by changes to \texttt{strings}. For a \texttt{string} fix, it is not always clear whether it really changes
the program behavior or has just a documenting function. This is in particular true when the fix changes a docstring which is still counted as a \texttt{string} token in Python. 
Therefore, by ignoring string fixes, we observe
that most operations modify function and method calls (e.g. "Insert argument list into call" or "Move identifier to attribute"). This is inline with our observation in Section 3 that the most frequent SStuBs address modifications of function calls.
Finally, to successfully repair a single statement bug, APR method should focus on operations that modify
tokens directly or extend program. Based on the high frequency of \texttt{Move} operations, a copy mechanism can
be a useful addition to an existing APR method. Fixing errors that are related to function calls are most helpful for a developer since they are most frequent bug fixes.

\subsection{How frequently do single statement bug fixes fix typos?}

\begin{table}
\caption{Percentage of single statement bugs that potentially occur due to a typo.}
\label{tab:edit-dist}
\centering
\begin{adjustbox}{min width=\columnwidth}
\begin{tabular}{l c c }
\toprule
{\bfseries Commit type} & {\bfseries \texttt{TSSB-3M}} &  {\bfseries \texttt{SSB-9M}} \\
\midrule
All single statement bugs & 22.84\% & 22.65\%\\
Change identifier used     & 30.50\% & 26.94\% \\
Change string used          & 42.10\% & 45.19\% \\
\bottomrule
\end{tabular}
\end{adjustbox}
\end{table}

An initial investigation revealed that single statement bugs occurring due to a programmer typo
are rather common (i.e. the developer forgets a bracket or misspelled an identifier). Therefore, we are interested
how frequent single statement bugs occur due to typos. For this, we assume that commits that modify not more than 
two characters (by inserting, removing, changing or transposing) represent a typo fix. To compute the number of text changes, we therefore employ the Damerau-Levenshtein edit distance~\cite{brill2000spell} between the code before and 
after the fix and count all changes with an edit distance smaller equal 2. Our results for \texttt{TSSB-3M} and \texttt{SSB-9M} are presented in Table \ref{tab:edit-dist}. We compute the percentage of single statement bugs per dataset that are considered to be a typo according to our heuristic. Additionally, we compute the same metric for all single statement bugs that only change a single identifier or string (since these are the most common single token fixes). To avoid counting single identifier fixes that completely replaces the identifier within two edit operations, we only count identifier that are at least 3 characters long. We apply the same rule for string fixes. 
In total, we found that typos are rather common which occur in at least 20\% of all our datasets. If we view only single identifier or string changes, the percentage even increases to over 30\% and over 40\% respectively. This indicates that are large portion of the single statement bug datasets are in fact likely related to typos. Therefore, it could 
be interesting direction for future research to explore APR methods that employ classical spelling correction techniques.

\end{document}